\documentclass[pdflatex,sn-mathphys-num]{sn-jnl}

\usepackage{graphicx}
\usepackage{amsmath,amssymb}
\usepackage{booktabs}
\usepackage{siunitx}
\usepackage{xcolor}
\graphicspath{{../figures/}}

\theoremstyle{thmstyleone}

\begin{document}

\title[MAGI: a generative ML framework for particle simulations]{MAGI
(Multivariate Autoencoder for particle Generative Inference): a generative Machine Learning
framework for accelerated multi-stage particle simulations}

\author*[1,2,3,4]{\fnm{Francesco} \sur{Monastra}\orcid{https://orcid.org/0009-0007-1675-6223}}\email{francesco.monastra@inaf.it}
\author[3]{\fnm{Marco} \sur{Pietrosanto}\orcid{https://orcid.org/0000-0001-5129-6065}}
\author[5]{\fnm{Giuseppe} \sur{Angora}\orcid{https://orcid.org/0000-0002-0316-6562}}
\author[5,6]{\fnm{Stefano} \sur{Cavuoti}\orcid{https://orcid.org/0000-0002-3787-4196}}
\author[3]{\fnm{Francesco} \sur{Fenu}\orcid{https://orcid.org/0000-0002-0313-8852}}
\author[1]{\fnm{Simone} \sur{Lotti}\orcid{https://orcid.org/0000-0002-3088-1561}}
\author[2,3]{\fnm{Matteo} \sur{Perri}\orcid{https://orcid.org/0000-0003-3613-4409}}
\affil[1]{\orgname{INAF-IAPS}, Via del Fosso del Cavaliere 100, 00133, Rome \country{Italy}}
\affil[2]{\orgname{INAF-OAR}, Via Frascati 33, 00078, Monte Porzio Catone (RM), \country{Italy}}
\affil[3]{\orgname{ASI-SSDC}, Via del Politecnico snc, 00133, Rome \country{Italy}}
\affil[4]{\orgname{University of Rome ``Tor Vergata''}, Via della Ricerca Scientifica 1, 00133, Rome, \country{Italy}}
\affil[5]{\orgname{INAF-OACN}, Salita Moiariello 16, 80137, Naples \country{Italy}}
\affil[6]{\orgname{INFN - Unit of Naples},Via Cinthia 6, 80126, Naples, \country{Italy} }

\abstract{Monte Carlo particle simulations are hindered by the computational cost of particle transport. This is especially true when large mass surrounds a small sensitive volume. MAGI is a conditional variational autoencoder that learns the joint energy, position and direction distribution of particles crossing a closed surface in a multi-stage simulation, to be resampled and re-injected in a smaller mass model. We validated it generating crossings, variables describing particle features later in the simulation, injecting them into the same Geant4 mass model and comparing the spectrum and counts at detector level with full simulation. Rates per injected crossing agree with unity at $\sim1\sigma$ on our benchmark models. We measure the gain of MAGI with statistical amplification $1/\chi$ and computational speed-up $S$: the first goes from 150 for high-statistics cosmic rays to 12000 for rare radioactive decays, while the latter ranges from 2.1 to almost 240, depending on source type and mass model. These first results validate MAGI v1.0.0 released on GitHub.}


\maketitle

\section{Introduction}\label{sec:intro}

Monte Carlo simulations using toolkits like Geant4 \cite{Agostinelli2003Geant4,Allison2006Geant4,Allison2016Geant4} are essential for background evaluation and sensitivity optimization in particle physics and astrophysics. However, in low-background experiments, particularly in cryogenic X-ray arrays with background requirements typically ranging from
$\sim10^{-2}$ \cite{Lotti2021XIFUBackground} down to
$\sim10^{-7}\,\mathrm{cts\,s^{-1}\,cm^{-2}\,keV^{-1}}$ \cite{Vaccaro2023}, particle transport through complex surrounding structures and shielding poses a critical computational bottleneck. High physical attenuation or low geometric acceptance means that most simulated primaries never interact with the active region, forcing redundant outer transport computations at every inner design iteration.

While generative surrogates emulate energy showers in fixed geometries \cite{Paganini2018CaloGAN,Krause2021CaloFlow} and phase-space interfaces like MCPL (Monte Carlo Particle Lists)\cite{Kittelmann2017MCPL} resample static recorded events without statistical amplification, MAGI (Multivariate Autoencoder for particle Generative Inference) bridges this gap. MAGI models the joint phase-space distribution of particles crossing a virtual surface enclosing the instrument. Decoupling the outer, design-invariant mass allows MAGI to rapidly generate synthetic crossings for downstream Geant4 re-injection, achieving both computational speed-up and statistical amplification.

\section{Method}\label{sec:method}

A virtual spherical shell encloses the detector assembly and records inward particle crossings in a primary Geant4 run. Geometry modifications confined within the shell can then be evaluated by generating synthetic crossings and transporting them inward.

MAGI is a conditional variational autoencoder (cVAE)\cite{Kingma2014VAE,Sohn2015CVAE} conditioned on particle species. A crossing consists of energy and four quantile-transformed directional/positional variables ($u_r, u_v, \phi_r, \phi_v$). The energy head combines a spline-flow continuum \cite{Durkan2019NeuralSplineFlows} with discrete components pinned at fluorescence lines under a focal objective \cite{Lin2017FocalLoss}. The latent prior uses a conditional coupling transform \cite{Dinh2017RealNVP,Tomczak2018VampPrior}. MAGI can also be conditioned on primary type and energy, preserving primary multiplicity.

\section{Validation}\label{sec:validation}

With $N_{\rm events}$ detector events from $N_{\rm crossings}$ injected crossings,
both arms, full Geant4 and MAGI re-injection, through the identical mass model, binary and scoring, we compare
\begin{equation}
R=\frac{N^{\rm MAGI}_{\rm events}/N^{\rm MAGI}_{\rm crossings}}
       {N^{\rm real}_{\rm events}/N^{\rm real}_{\rm crossings}} .
\label{eq:R}
\end{equation}
At $R=1$ a MAGI crossing yields events at the real rate. The
crossing efficiency $\chi=N^{\rm real}_{\rm crossings}/N^{\rm real}_{\rm primaries}$
is the geometrical and physical attenuation of the outer mass. The statistical amplification
$\varepsilon=R/\chi$ is the gain in statistics per generated crossing, and an ideal
surrogate returns $1/\chi$.

\begin{figure}[t]
\centering
\includegraphics[width=0.85\textwidth]{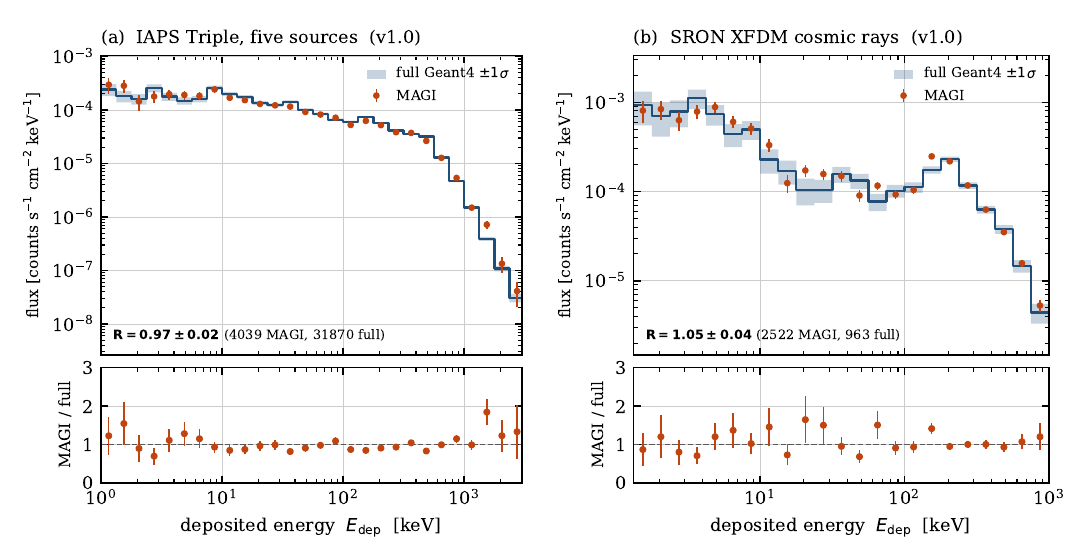}
\caption{Detector response, MAGI against full Geant4 in the same mass model, with
$\pm1\sigma$ Poisson bands and MAGI-to-full ratios in the bottom panels. (a) IAPS Triple, v1.0, five sources ($^{232}$Th, $^{226}$Ra, $^{40}$K, $\gamma$, $\mu$). (b) SRON XFDM cosmic rays, v1.0, against
$2.4\times10^{6}$ re-injected real crossings.}
\label{fig:spectra}
\end{figure}

On first benchmark model (IAPS Triple) the real arm is a run of $2.7\times10^{9}$ to $5.4\times10^{10}$ primaries per source, performed on our cluster (Dual socket Xeon 8268, 96 cores per 6 nodes, 384GiB, 540 threads)  and the MAGI arm on a MacBookAir (AppleM1, 4 performance and 4 efficiency cores, 16GB) with $8\times10^{5}$ to $4\times10^{6}$ crossings per source. The ratio obtained for this model is
$0.97\pm0.02$  (Fig.~\ref{fig:spectra}a) over the sum of all sources. A second benchmark model (SRON XFDM)  gives $R=1.01\pm0.03$ over all deposits (2550 against
2565 events, $6.0\times10^{6}$ crossings per arm) and $1.05\pm0.04$ in 1\,keV to 1\,MeV, plotted
(Fig.~\ref{fig:spectra}b). 
Our validation is performed with Geant4 10.4.3.

\section{Amplification and speed-up}\label{sec:cost}

Amplification $1/\chi$ measures detector events
gained per generated crossing instead of per thrown primary. Speed-up $S$ compares
computational cost: a crossing starts inside the assembly and costs more to transport than a
primary, so
\begin{equation}
S=\frac{1}{\chi}\,\frac{c_{\rm primary}}{c_{\rm crossing}},
\label{eq:S}
\end{equation}
with both costs in core-hours, $c_{\rm primary}$ the whole-run cost per primary and
$c_{\rm crossing}$ the inner-transport cost per injected crossing. The speed-up $S$ of Eq.~\ref{eq:S}
omits the outer campaign that records the training set, and training time and represents a theoretical limit given a specific configuration. Let $N_0$ be the primaries behind the training set, $C_{\rm outer}=N_0c_{\rm primary}$ the cost of recording it, $C_{\rm train}$ the training cost, and $c_{\rm gen}$ the generation cost of the re-injected particle. For a simulation target of $A\times N_0$ primaries, the net speed-up
is the full cost over the MAGI cost,

\begin{equation}
S_{\rm net}(A)=\frac{A\,C_{\rm outer}}
{C_{\rm outer}+C_{\rm train}+C_{\rm re-inj}},
\label{eq:Snet}
\end{equation}

where $C_{\rm re-inj} = A\,C_{\rm outer}\,\chi\,\frac{(c_{\rm gen}+c_{\rm crossing})}{c_{\rm primary}}$ is the re-injection cost.
$S_{\rm net}(A)$ tends to $S$ for large $A$ and sits just below 1 at $A=1$.

Table~\ref{tab:speedup} reports $^{232}$Th, $^{226}$Ra, $^{40}$K and cosmic rays (CR, $\gamma$ and $\mu$)  in the two tested models, on the laptop, using MAGI v1.0 and $6$ parallel jobs per arm. Generating $10^{6}$ crossings takes $46$\,s on one
thread ($c_{\rm gen}=1.27\times10^{-8}$ core-h per crossing). Training on the ASI-SSDC Hub CPU (Xeon Gold 6252N, 2.3 GHz, 96 cores, 128 GiB, 12 threads) takes $2.9\times10^{3}$ to $7.3\times10^{3}$\,s for 40 epochs plus the prior refit, that is $10$ to $24$ core-h. 

\begin{table*}[t]
\caption{Amplification and speed-up on two mass models, $c$ in core-h. $A$ is in units of $N_0$.}
\label{tab:speedup}
\centering\footnotesize
\setlength{\tabcolsep}{4pt}
\begin{tabular}{@{}lcccc|ccccc@{}}
\toprule
 & \multicolumn{4}{c|}{SRON XFDM} & \multicolumn{5}{c}{IAPS Triple} \\
 & $^{232}$Th & $^{226}$Ra & $^{40}$K & CR & $^{232}$Th & $^{226}$Ra & $^{40}$K & $\gamma$ & $\mu$ \\
\midrule
$N_0$ ($10^{9}$)               & 3.6 & 3.6 & 36 & 0.54 & 5.4 & 5.4 & 54 & 2.7 & 1.8 \\
$1/\chi$ (amplification)       & 464 & 601 & 6542 & 144.5 & 668 & 869 & 12175 & 276 & 153 \\
$c_{\rm primary}$ ($10^{-8}$)  & 10.1 & 7.83 & 1.28 & 28.6 & 16.8 & 14.3 & 2.54 & 12.0 & 62.6 \\
$c_{\rm crossing}$ ($10^{-7}$) & 3.31 & 3.63 & 3.39 & 194 & 22.6 & 20.3 & 25.3 & 30.6 & 232 \\
$S$, Eq.~\ref{eq:S}            & 142 & 130 & 247 & 2.1 & 50 & 61 & 122 & 10.8 & 4.1 \\
$S_{\rm net}$, $A=10$          & 8.8 & 8.6 & 9.1 & 1.7 & 8.1 & 8.4 & 9.1 & 5.0 & 2.9 \\
$S_{\rm net}$, $A=100$         & 56 & 53 & 68 & 2.1 & 33 & 37 & 54 & 9.6 & 4.0 \\
$S_{\rm net}$, $A=10^3$        & 119 & 110 & 190 & 2.1 & 47 & 57 & 108 & 10.6 & 4.1 \\
$S_{\rm net}$, $A\to\infty$    & 137 & 125 & 238 & 2.1 & 49 & 61 & 122 & 10.8 & 4.1 \\
\bottomrule
\end{tabular}
\end{table*}

The computational gain strictly follows the source rarity. The heavily attenuated $^{40}$K achieves the highest performance, reaching $S=247$ (SRON XFDM) and $122$ (IAPS Triple), with net values of $190$ and $108$ at $A=10^3$, followed by the intermediate radioisotopes $^{232}$Th and $^{226}$Ra. This behavior reflects the balance between physical attenuation $1/\chi$ and computational cost: inner transport per injected crossing ($c_{\rm crossing}$) is intrinsically more expensive than outer primary transport ($c_{\rm primary}$). 
Conversely, high-statistics and penetrating sources gain the least: gammas, muons, and cosmic rays require significantly higher transport cost per crossing than radiosources, yielding more modest speed-ups of $S=10.8$, $4.1$, and $2.1$, respectively. All reported metrics represent single-laptop benchmarks.

\section{Limitations}\label{sec:limits}

MAGI v1.0 exhibits structural limits arising from cVAE latent capacity bounds (prior-posterior mismatch) and the conditional independence of decoding heads given $z$. While localized spectral resolution defects (such as fluorescence line smearing) are mitigated in v1.0 via explicit delta-line sampling, fundamental head decorrelation and shower-context conservation cannot be resolved without structural changes. The implementation of autoregressive prior flows, coupled flow heads, and context-aware shower conditioning is deferred to the MAGI v2.0 architecture in development.

\section{Conclusions}

MAGI reproduces the detector response of a full Geant4 run within $3\,\%$
($R=0.97\pm0.02$, five sources pooled) on IAPS Triple and on SRON XFDM ($R=1.05\pm0.04$). On a laptop it amplifies statistics by 464 to 12175 and speeds up the computation by 50 to 247 for radioactive sources depending on the mass model.
This tool improves the efficiency and the affordability of particle simulations, demonstrating also the feasibility of
running the same setup on a laptop instead of a dedicated cluster.
The architecture and the user manual are available on GitHub with full package and examples.

\noindent\textbf{Declarations.} \emph{Licence:} GPL v3.0,
\texttt{github.com/francescomonastra/MAGI}. \emph{Funding:} ASI-INAF agreement
n.\ 2025-33-HH.0. \emph{Use of Generative AI:} During the preparation of this work, the authors used generative AI model (Claude by Anthropic) for language polishing, text editing, and LaTeX formatting support. The authors reviewed and edited all content as needed and take full responsibility for the final text and data. \emph{Competing interests:} none.

\bibliography{references}

\end{document}